\documentstyle[12pt,epsf]{article}
\let\section=\subsection     \let\subsection=\subsubsection                
\begin{document}
\newcommand{\p}{\partial}
\newcommand{\ls}{\left(}
\newcommand{\rs}{\right)}
\newcommand{\beq}{\begin{equation}}
\newcommand{\eeq}{\end{equation}}
\newcommand{\beqa}{\begin{eqnarray}}
\newcommand{\eeqa}{\end{eqnarray}}
\newcommand{\beqao}{\begin{eqnarray*}}
\newcommand{\eeqao}{\end{eqnarray*}}
\newcommand{\bdm}{\begin{displaymath}}
\newcommand{\edm}{\end{displaymath}}
\begin{center}
   {\large \bf TREATMENT OF BARYONIC RESONANCES IN}\\[2mm]
   {\large \bf THE RQMD APPROACH INCLUDING }\\[2mm]
   {\large \bf SCALAR--VECTOR MEAN FIELDS}\\[5mm]
   C. FUCHS, L. SEHN, AMAND FAESSLER, D.S. KOSOV, V.S. UMA MAHESWARI 
and Z. WANG \\[5mm]
   {\small \it  Institut f\"ur Theoretische Physik, Universit\"at T\"ubingen,\\
Auf der Morgenstelle 14, D-72076 T\"ubingen, Germany \\[8mm] }
\end{center}

\begin{abstract}\noindent
In the relativistic Quantum Molecular Dynamics (RQMD) approach 
baryons are described within the framework of covariant 
hamilton constraint dynamics. The inclusion of a relativistic mean 
field results in a quasiparticle picture for the 
baryons. This requires to distinguish between canonical 
and kinetic variables of the particles. 
As resonances we include the $\Delta$(1232) and the 
$N^*$(1440) resonance. The resonance masses are 
distributed according Breit-Wigner functions. 
However, the scalar self energy 
leads to a shift in the masses and introduces an additional medium 
dependence. Consequences of this 
description on resonance and pion dynamics are discussed.
\end{abstract}

\section{Introduction}
In relativistic heavy ion collisions, e.g. in the SIS energy range 
excited nucleon states play a decisive role in the reaction dynamics. 
Most important is the $\Delta$(1232) resonance which can reach 
abundances comparable to nuclear matter saturation density \cite{metag93}, 
however also the $N^*$(1440) plays a non-negligible role. The decay 
and rescattering of these resonances are the predominant sources of 
meson production. 
In order to use such mesons as a source of information of the hot 
and compressed phase in heavy ion reactions the understanding of their 
origin, i.e. of the properties of excited nucleon states appear to 
be indispensable. 
Microscopic studies of the excitation and deexcitation of nucleons can be 
performed within the framework of transport models like
Quantum Molecular Dynamics (QMD) \cite{Aichelin91}
and the BUU-type models. 
 
In the covariant extension of QMD, i.e., the Relativistic QMD (RQMD)
\cite{Sorge89,Lehmann95} up to now only
static Skyrme forces were used. These generalized Skyrme forces 
are treated as scalar potentials in the framework 
of Constrained Hamilton Dynamics. But the full
Lorentz-structure of the nucleon-nucleon interaction contains large scalar and 
vector components. Effects of these relativistic forces 
were already studied in the
framework of covariant generalizations of models of the BUU-type
\cite{Ko89,Cassing90b,fuchs95}, mostly in the framework of the 
Walecka model \cite{Serot88} and its non-linear extensions, 
but not yet in RQMD. In Ref. \cite{fuchs96b} the formulation 
of a sclar-vector RQMD has been given. Here we study in particular the 
influence of the relativistic 
mean field on the resonance dynamics.  

\section{RQMD with scalar-vector mean fields}
The relativistic self-energy $\Sigma = \Sigma_s - \gamma_\mu \Sigma^\mu$ 
contains scalar and vector components. 
Thus, one has to distinguish between canonical momenta $p_i$ and bare 
masses $M_i$ on the one hand and kinetic momenta $p^{*}_i$ and 
effective (Dirac) masses $m^{*}_i$ on the other hand. 
The latter correspond to the quasiparticles dressed by the surrounding medium 
which obey the in-medium Dirac equation 
\beq
\left( \gamma_\mu p^{*\mu}_i - m^{*}_i \right) u^{*}_i = 0
\label{dirac}
\eeq
where $u^{*}_i$ is an in-medium spinor or a Rarita spinor in the case of 
a $\Delta$. 

In the formalism of Constrained Hamiltonian Dynamics \cite{Dirac49,Samuel82}, 
the 8N dimensional phase space of 
N interacting relativistic particles is reduced 
to 6N dimensions by 2N
constraints which fix the individual energies and 
by the mass-shell constraints parametrize the world 
lines by fixing the relative time 
coordinates.  

With respect to Eq. (\ref{dirac}) the N on-mass-shell 
conditions for the four-momenta are 
given in terms of the quasiparticles 
\beqa
K_{i} = (p^{\mu}_{i}-\Sigma^{\mu}_{i})(p_{i\mu}-\Sigma_{i\mu})
-(M_{i}-\Sigma_{is})^{2} 
= p^{*}_{i\mu} p^{*\mu}_{i} - m^{*2}_i = 0 
\quad ,
\label{const1}
\eeqa
which contain scalar and vector self-energies. Hence, the constraints, 
Eq.(\ref{const1}), take the full 
Lorentz structure of the relativistic mean field into account. 
In the relativistic Hartree approximation the scalar and vector 
self-energies are proportional 
to the scalar density 
$\rho_{s}$ and the baryon current $B^{\mu}$, respectively
\beqa
\Sigma_s = \Gamma_{s} \rho_{s} \quad ,\quad 
\Sigma^{\mu} = \Gamma_{v}B^{\mu}
\quad .
\label{sigma1}
\eeqa
The proportionality factors $\Gamma_{s,v}$ in Eq. (\ref{sigma1}) 
may either be constants as in the Walecka model \cite{Serot88} or, 
if one includes higher order medium effects, density dependent functions as 
discussed, e.g, in Ref. \cite{fuchs96b}. Here we apply the 
non-linear Walecka model (NL2) which can be expressed in the form 
of Eq. (\ref{sigma1}) with   
\beq
\Gamma_s (\rho_s) = \frac{g_{\sigma}^2}
{ m_{\sigma}^2 + B\Phi(\rho_s) + C\Phi^2(\rho_s) }  
\eeq
where $\Phi$ is the scalar $\sigma$--meson field \cite{fuchs96b}. 
The scalar density and the baryonic current 
\beqa 
\rho_{s}(q_{i},p_{i})=\sum_{j \neq i}\rho_{ij} 
\quad , \quad 
B_{\mu} (q_i,p_{i})=\sum_{j \neq i}\rho_{ij} u_{j\mu}
\label{dens}
\eeqa
are determined in terms of the scalar two-body densities
\begin{equation}
\rho_{ij}=\frac{1}{(4\pi L)^{3/2}}exp(q^{2}_{Tij}/4L) \quad ,
\end{equation}
with $u^{\mu}_{j} = p^{*\mu}_{j} / m^{*}_j $ being the 4-velocity
of particle $j$ and $q_{Tij}$ the invariant center-of-mass distance 
of the particels $i$ and $j$ as defined in Ref. \cite{Sorge89}. 

The total Hamiltonian is given by the N on-shell constraints, Eq. 
(\ref{const1}), and by N-1 time fixation constraints $\phi_i$ choosen as 
in \cite{Sorge89}. These time constraints ensure that interacting 
particles have equal times in their center-of-mass frame. 
A final constraint fixes the global time evolution 
parameter. The Hamiltonian  
\begin{equation}
H=\sum_{i=1}^{N}\lambda_{i} K_i + \sum_{i=N+1}^{2N-1}\lambda_{i} \phi_{i}
\label{hamilton}
\end{equation}
generates the equations of motions for the canonical conjugated coordinates 
and momenta  
\begin{equation}
dq^{\mu}_{i}/d\tau=[H,q^{\mu}_{i}]
\quad , \quad dp^{\mu}_{i}/d\tau=[H,p^{\mu}_{i}]
\label{rqmd1}
\end{equation}
where $[A,B]$ means the Poisson bracket of phase space functions $A$ and $B$. 
Notice that in the present formalism canonical and kinetic momenta are 
no independent quantities. 
To integrate the set of above equations (\ref{rqmd1}), 
one has to determine the unknown Lagrange multipliers 
$\lambda_{i}(\tau)$. This can be done using the fact that the complete 
set of $2N$ constraints must be fulfilled during the 
whole time evolution, i.e. $d K_{i}/d\tau = d\phi_{i}/d\tau= 0$. 
If the Dirac's first class condition is fulfilled, i.e. $[K_{i},K_{j}]=0$, 
the Hamiltonian is reduced to
\beqa
H=\sum_{i=1}^{N}\lambda_{i}K_{i} \quad ,\quad 
\lambda_{i}=\Delta^{-1}_{iN}
\eeqa
where $\Delta_{ji}=[K_{j},\chi_{i}]$ is a submatrix 
of the complete constraint matrix $C_{ji}$ \cite{Sorge89,Lehmann95}.

In this approach the dynamics are to high extent determined by the 
Lagrange multipliers $\lambda_{i}(\tau)$. Thus in Fig.1 we show the 
time evolution of $\lambda$ for a representative nucleon in a central 
Ca+Ca collision at 1.85 A.GeV for both, a pure mean field calculation and 
including binary collisions. It is clearly seen that $\lambda$ is a smooth 
function in time if its evolution is governed only by the mean field dynamics. 
However, in the high density phase ($10 fm/c \leq t \leq 20 fm/c $) 
strong fluctuations occure which are due to binary collisions. These 
fluctuations represent rapid changes in the momenta and -- 
in the case of a resonance creation or decay -- the masses of the 
particles.

\begin{minipage}{13cm}
\begin{center}
\leavevmode
\epsfxsize=10cm
\epsffile[60 210 500 520]{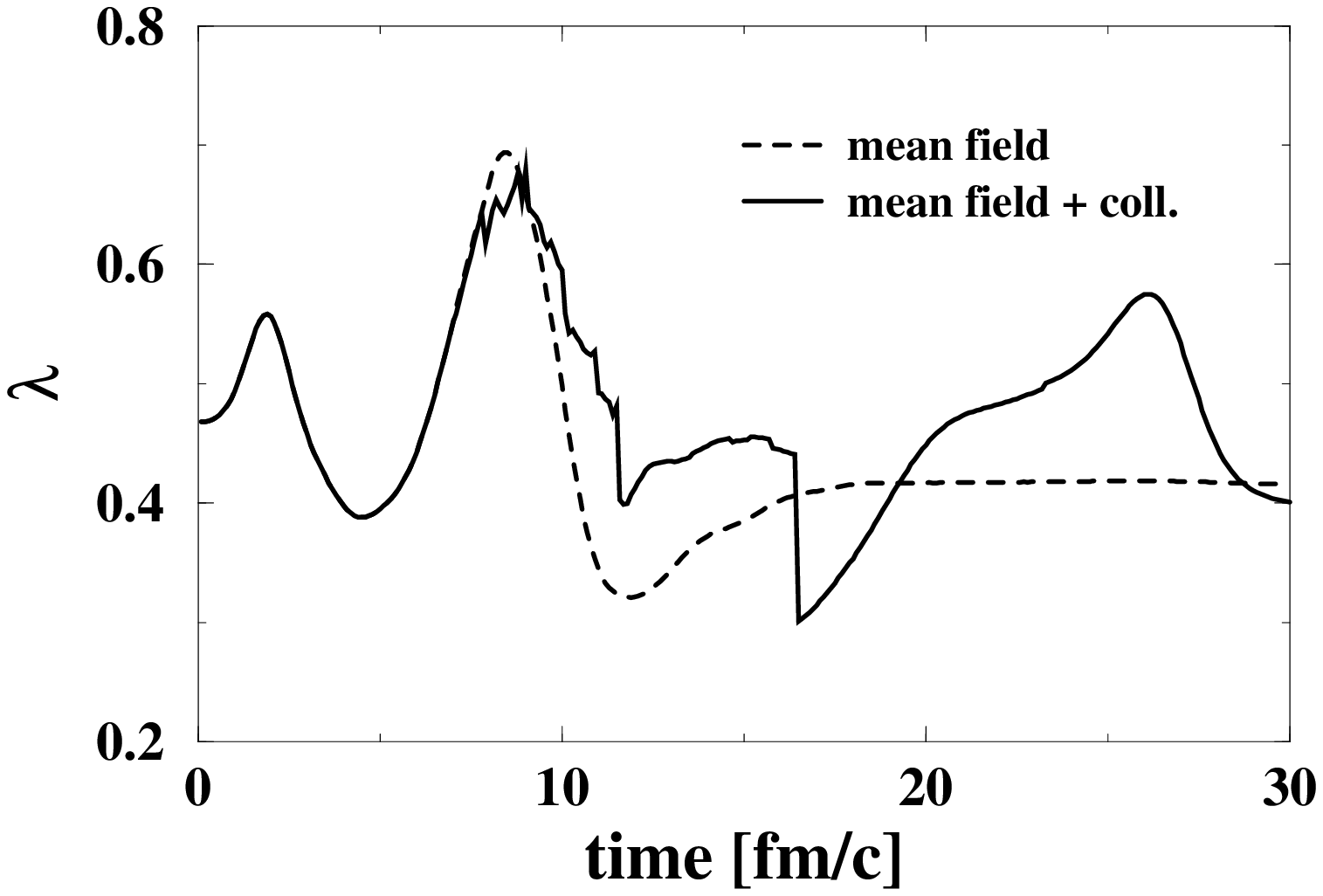}
\end{center}
{\begin{small}
Fig.~1. Time evolution of a lagrange multiplier in a central Ca+Ca reaction 
at 1.85 A.GeV. A full calculation including binary collisions (solid) is 
compared to a pure mean field calculation (dashed).
\end{small}}
\end{minipage} \\ \\ \\
\noindent
The inclusion of a relativistic mean field which introduces 
an additional momentum dependence into the model is essential 
in order to retain the corrcet dynamics at incident energies above 
about 1 A.GeV. In Fig.2 we compare the present approach to the 
standard RQMD \cite{Sorge89,Lehmann95} which uses static Skyrme forces. 
From Fig.2 which shows the transverse flow per particle 
for the reaction Ar+KCL at 1.8 A.GeV under minimum bias condition 
it is clearly seen that the Skyrme forces result in far too less 
repulsive mean fields whereas NL2 is in good agreement with the data.
\section{Baryonic resonances}
The mean field propagation described in the previous section 
is the same for nucleons and baryonic resonances. 
In the latter case the bare nucleon mass 
in Eq. (\ref{const1}) has to be replaced by the corresponding 
resonance mass $M_R$, i.e. $m^{*}_R = M_R - \Sigma_{sR}$. Thus we 
apply the {\it same} mean field to nucleons and resonances, i.e., the 
coupling strenght of the respective mesons is assumed to be identical. 
Since we restrict to non-strange mesons this appears to be reasonable 
In the present approach the coupling 
functions $\Gamma_{s,v}$ have to be interpreted 
as effective quantities which parametrize the mean field rather 
than elementary vertex functions which further justifies 
this assumption. 

\begin{minipage}{13cm}
\begin{center}
\leavevmode
\epsfxsize=10cm
\epsffile[0 100 420 390]{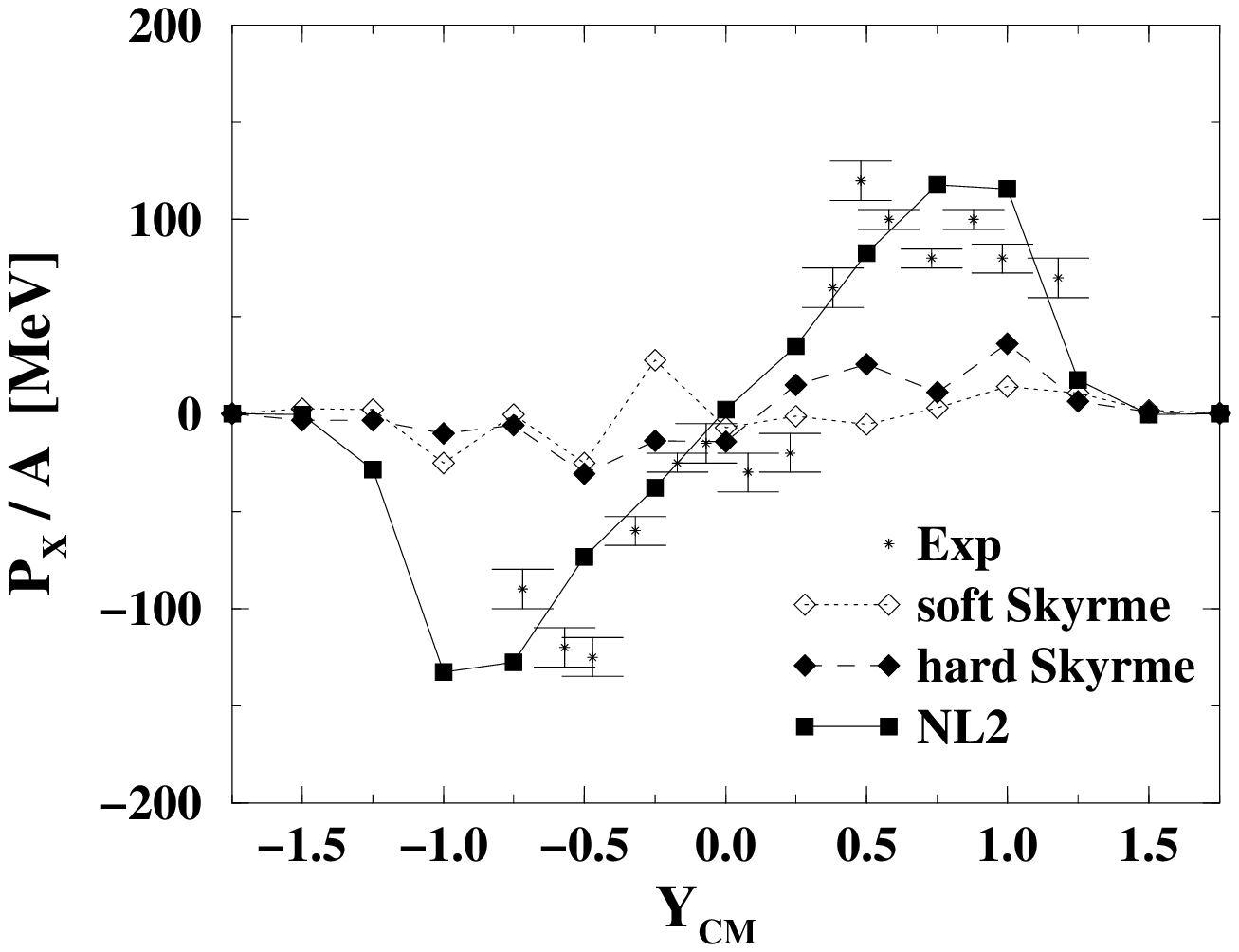}
\end{center}
{\begin{small}
Fig.~2. Transverse flow per particle for the reaction Ar+KCl 
at 1.8 A.GeV. Static Skyrme forces (soft/hard) are compared to the non-linear 
Walecka model NL2 and to data from Ref. \cite{Danielewicz85}.
\end{small}}
\end{minipage}\\ \\ 
\noindent
For the inelastic 
nucleon-nucleon channels we include the $\Delta(1232)$ as well as 
the $N^{*}(1440)$ resonance with the cross sections of Ref. \cite{Hu94}. 
The lifetimes of the resonances are 
determined through their energy and momentum dependent decay widths 
\beq
\Gamma (|{\bf p}|) = \frac{a_1 |{\bf p}|^3}
{(1+ a_2 |{\bf p}|^2 )(a_3 + |{\bf p}|^2) } \Gamma_0
\label{reswidth}
\eeq
which originates from the $p$-wave representation of the 
resonances. In Eq. (\ref{reswidth}) ${\bf p}$ is the momentum of the 
created pion (in GeV/c) in the resonance rest frame. 
According to Ref. \cite{Hu94} the values 
$a_1$=22.83 (28.8), $a_2$=39.7 and $a_3$=0.04 (0.09) are used for 
the $\Delta$ ($N^*$) and the bare decay widths are taken as  
$\Gamma_{0}^\Delta$= 120 MeV and $\Gamma_{0}^{N^*}$= 200 MeV. 

Since only the quasiparticles, i.e. kinetic momenta and 
effective mass lie on the mass-shell all 
collisions are performed in the kinetic center-of-mass frame with 
$\sqrt{s^*} = \sqrt{ {\bf p}^{* 2}_{CM} + m^{* 2}_{1} } + 
\sqrt{ {\bf p}^{* 2}_{CM} + m^{* 2}_{2} }$. In the case of an 
inelastic collision, e.g. $NN\longmapsto NR$ the final 
momentum can be evaluated 
\beq
|{\bf p}^{*}_{CM}| = \frac{1}{2\sqrt{s^*}}
\sqrt{\lambda (s^*, m^{*}_N , m^{*}_R)}
\eeq
with 
$ \lambda = [ s^* -(m^{*}_N +m^{*}_R)^2 ][ s^* -(m^{*}_N -m^{*}_R)^2 ]$. 
The probability distribution for the effective resonance mass is given 
by a Breit-Wigner distribution 
\beq
A(m^{*}_R) = \frac{1}{\pi} 
\frac{ \Gamma / 2}{ (m^{*}_R - m^{0*}_R )^2 + (\Gamma / 2)^2 }
\eeq
with $m^{0*}_R = M^{0}_R - \Sigma_{sR}$. The maximal possible bare 
mass of the created resonance is then restricted by 
\beq
M_{R}^{max} = \sqrt{s^*} - m^{*}_N + \Sigma_{sR} 
\quad .
\eeq
\begin{minipage}{13cm}
\begin{center}
\leavevmode
\epsfxsize=10cm
\epsffile[30 90 450 400]{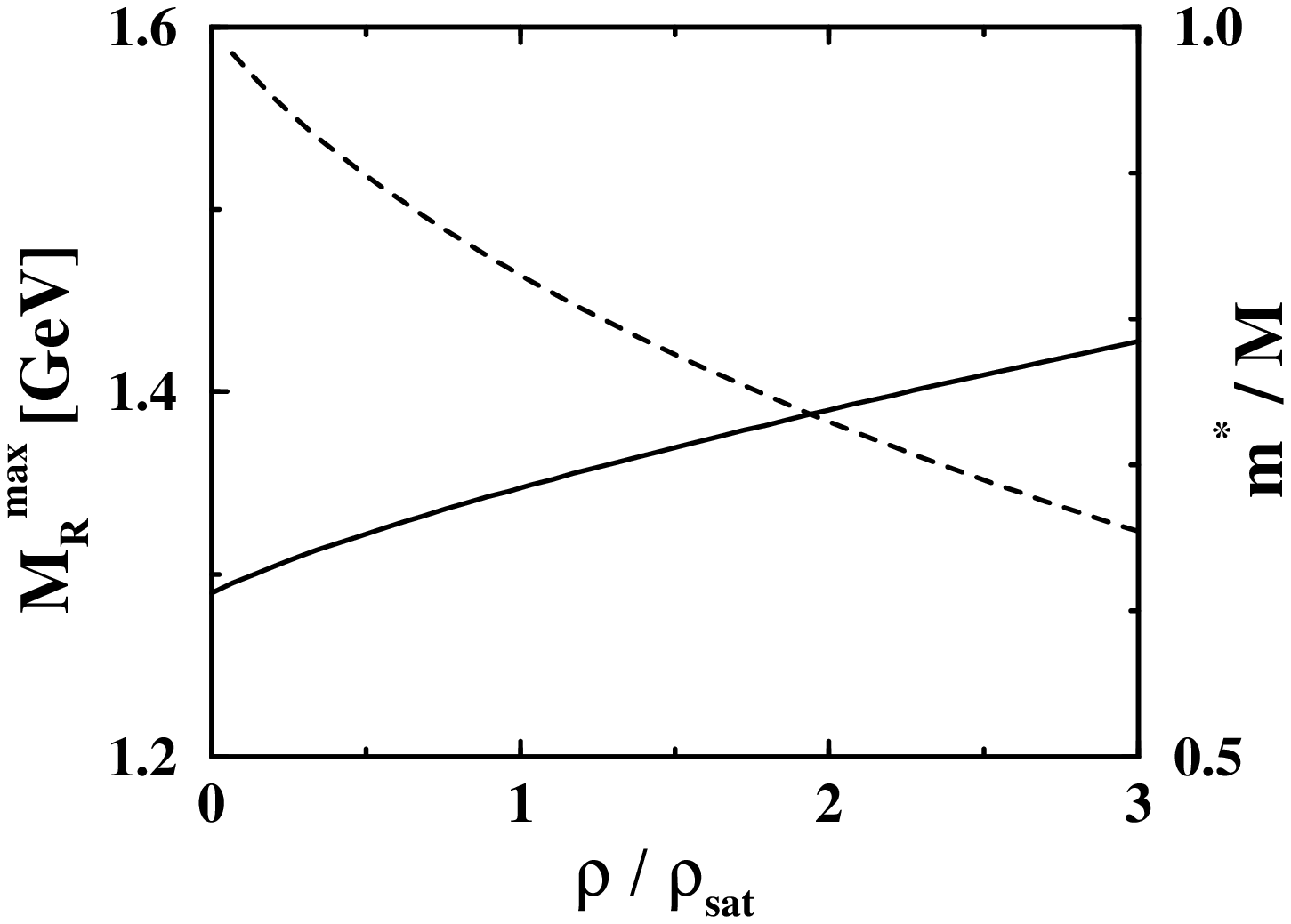}
\end{center}
{\begin{small}
Fig.~3. Shift of the maximal resonance mass by the scalar 
self-energy in nuclear matter (solid, left scale). In addition the 
effective nucleon mass $m^*$ is shown (dashed, right scale). The non-linear 
Walecka model NL2 is applied.
\end{small}}
\end{minipage}\\ \\ \\ 
\noindent
Fig.3 shows the medium dependence of $M_{R}^{max}$ created in a 
$NN\longmapsto NR$ collision. Here we have choosen the momenta of 
the incident nuclei (${\bf p}^{*}_1 = - {\bf p}^{*}_2$) to be 0.6 GeV/c 
in the nuclear matter rest frame. It is seen that $M_{R}^{max}$ 
is considerably enhanced by the presence of 
the medium, i.e. by the attractive 
scalar self-energies which is correlated to the reduction of the 
effective mass $m^*$ also shown in Fig.3. Thus the probability for 
the excitation of higher resonances is generally enhanced in 
the presence of relativistic mean fields. This effect is even more 
pronounced when models are applied which result in larger fields than 
NL2 as, e.g., the original Walecka model \cite{Serot88}.  

The reabsorption processes ($\pi N\rightarrow \Delta,N^{*}$) are 
treated as described in Ref. \cite{fuchs97} again adopting the 
cross sections from Ref. \cite{Hu94}. 

Finally in Fig.4 we show the pion $p_t$-spectrum obtained in a 
1.85 A.GeV Ni+Ni collision. The contributions from pions originating from 
$N^*$--resonances are shown seperately and it is seen that these pions 
contribute over all by about 10\% to the total yield. However, 
$N^*$ pions are in particular relevant for the high energetic part 
of the spectrum. Here the contributions from $N^*$ are more than 20\%. 
Since high energetic pions are supposed to be most likely produced 
in the early phase of the reaction \cite{bass} they really probe 
the high density phase and thus can give signals on the medium 
dependence of the resonances.

\begin{minipage}{13cm}
\begin{center}
\leavevmode
\epsfxsize=11cm
\epsffile[20 85 430 400]{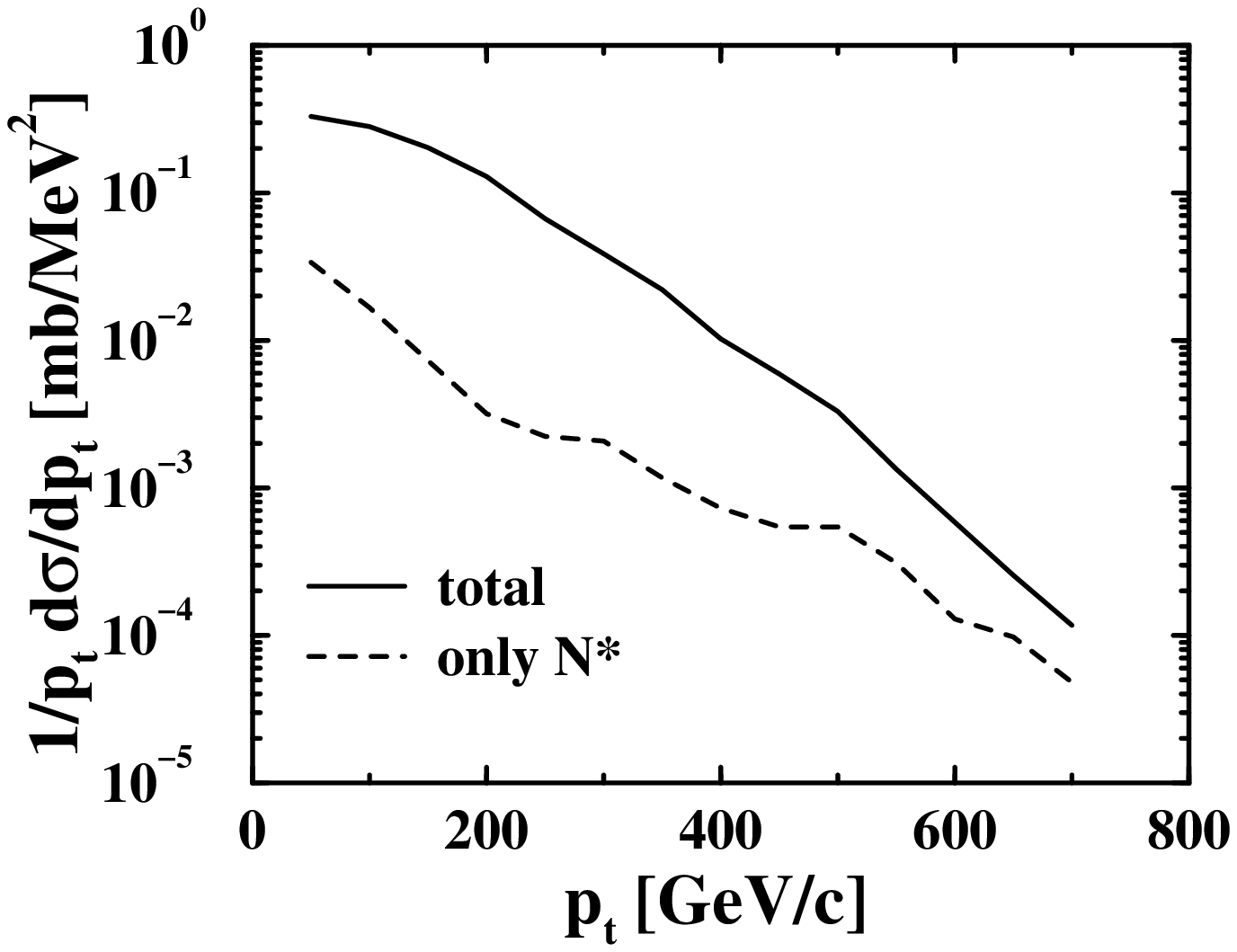}
\end{center}
{\begin{small}
Fig.~4. Transverse $\pi^0$ spectrum in a Ni+Ni collision at 
1.85 A.GeV under minimal bias condition. The contribution from 
pions originating from $N^*$ are shown seperately.
\end{small}}
\end{minipage}\\ \\ \\ 
\noindent
To summarize we have applied relativistic mean fields with 
scalar and vector components in the formalism of Hamilton Contraint 
Dynamics. This results in a quasiparticle picture for baryons and 
resonances. The attractive scalar fields lead thereby to a shift 
of the resonance mass distributions towards higher values and thus 
reduces the thresholds for the excitation of high lying resonances. 
Concerning heavy ion reactions in the SIS domain 
we found that the $N^*$ gives important contributions to the pion spectra. 
In particular high energetic pions will be well adopted to 
study the influence of medium effects on baryonic resonances.


\end{document}